\documentclass[twocolumn]{aastex62}
\usepackage{amsmath}
\usepackage[switch]{lineno}

\received{, 2023}
\revised{, 2023}
\accepted{, 2023}

\submitjournal{ApJ}

\shorttitle{Multi-wavelength analysis of GRB 171205A}
\shortauthors{Li et al.}

\begin{document}

\title{Multi-wavelength analysis of the SN-associated low-luminosity GRB 171205A}
\author[0000-0001-6469-8725]{Li, Xiu-Juan}
\affiliation{School of Physics and Physical Engineering, Qufu Normal University, Qufu 273165, China, z\underline{~}b\underline{~}zhang@sina.com}
\affiliation{School of Cyber Science and Engineering, Qufu Normal University, Qufu 273165, China, lxj@qfnu.edu.cn}
\author[0000-0001-6469-8725]{Zhang, Zhi-Bin$^{\dag}$}
\affiliation{School of Physics and Physical Engineering, Qufu Normal University, Qufu 273165, China, z\underline{~}b\underline{~}zhang@sina.com}
\author[0000-0001-7199-2906]{Huang, Yong-Feng}
\affiliation{School of Astronomy and Space Science, Nanjing University, Nanjing 210023, China}
\affiliation{Key Laboratory of Modern Astronomy and Astrophysics (Nanjing University), Ministry of Education, Nanjing 210023, China}
\affiliation{Xinjiang Astronomical Observatory, Chinese Academy of Sciences, 150 Science 1-Street, Urumqi 830011, China}
\author[0000-0001-7943-4685]{Xu, Fan}
\affiliation{School of Astronomy and Space Science, Nanjing University, Nanjing 210023, China}

\begin{abstract}
Multi-wavelength properties of the nearby Supernova(SN)-associated low-luminosity GRB
171205A are investigated in depth to constrain its physical
origin synthetically. The pulse width is found to be correlated
with energy with a power-law index of $-0.24\pm0.07 $, which is consistent
with the indices of other SN/GRBs but larger than those of long GRBs. By analyzing the overall light curve of its prompt
gamma-rays and X-ray plateaus simultaneously, we infer that the early X-rays together with the gamma-rays
should reflect the activities of central engine while the late X-rays may be dominated by the interaction of external shocks
with circumburst material. In
addition, we find that the host radio flux and offset of GRB 171205A are similar to
those of other nearby low-luminosity GRBs. We adopt 9 SN/GRBs with measured offset to
build a relation between peak luminosity ($L_{\gamma,p}$) and spectral lag ($\tau$) as $L_{\gamma,p}\propto\tau^{-1.91\pm0.33}$. The peak luminosity and the projected physical offset of both 12 SN/GRBs and 10
KN/GRBs are found to be moderately correlated, suggesting their different progenitors.  The multi-wavelength afterglow fitted with a top-hat jet model indicates that the jet half-opening angle and the viewing angle of GRB 171205A are $\thicksim$ 34.4 and 41.8 degrees, respectively, which implies that the off-axis emissions are dominated by the peripheral cocoon rather than the jet core.
\end{abstract}
\keywords{
\href{http://astrothesaurus.org/uat/739}{High energy astrophysics (739)};
\href{http://astrothesaurus.org/uat/629}{Gamma-ray bursts (629)};
}

\section{Introduction} \label{sec:intro}
Gamma-ray bursts (GRBs) are the most energetic gamma-ray flashes
in the universe. They are consisted of two stages: an initial
prompt gamma-ray emission phase and a long-lasting multi-wavelength afterglow emission phase
\citep{Zhang2004,ZhangB2007,Kumar2015,Zhangb2018}. Physically, the prompt emissions are
considered to originate from internal shocks due to the
interactions between ejected materials inside the fireball, while
the broadband afterglows are produced by external shocks due to
ejecta-medium interactions
 \citep[e.g.][]{Meszros1997, Sari1997, Huang2000,Zhangb2018}.
In general, the collapsar model is thought to account for long
GRBs (LGRBs) with a duration of $T_{\rm 90} >$ 2 s
\citep{Paczy1998,Woosley2006}. So far, a handful of LGRBs are
observed to be associated with energetic, broad-lined,
stripped-envelope supernovae (SNe). These SN-associated GRBs
usually have a luminosity 3 $\thicksim$ 4 orders of magnitude
lower than normal LGRBs \citep[e.g.][]{Galama1998,Pian2006}.

The light curves of prompt emissions are variable and highly
irregular, reflecting the temporal properties of the internal
energy dissipation and the activities of the cental engine
\citep[e.g.][]{Rees1994,Lxj1}. Besides, the spectral properties of
GRBs contain the key information of radiative mechanisms
\citep{Norris1996}. For example, the time lags of light curve
between different energy bands can be used to understand the
emission mechanism of prompt emissions
\citep{Zhangzb2006a,Wei2017}. Generally, when the higher energy
photons arrive earlier than lower energy photons, we call it a
positive time lag. Otherwise, it is called a negative lag.
\cite{Norris2000} reported an anti-correlation between the peak
luminosity ($L_{\rm \gamma,p}$) and the time lag ($\tau$) for six
BATSE LGRBs, which can be described by a power-law function as
$L_{\rm \gamma,p} \propto \tau^{\rm -1.14}$. At the same time,
they found that the SN-associated GRB 980425 falls below the
extrapolated power-law function by a factor of 400 -- 700.
Subsequently, the anti-correlation between
$L_{\rm \gamma,p}$ and $\tau$  was further investigated and
confirmed with different samples \cite[e.g.][]{Norris2002,Schaefer2007,
Hakkila2008,Arimoto2010,Ukwatta2010,
Ukwatta2012,Bernardini2015}. However, whether the $L_{\rm \gamma,p}-\tau$ relation exists for the SN/GRBs is unknown yet until
\cite{Li2023} recently  built the luminosity
relation $L_{\gamma,p}\propto \tau^{-1.43\pm0.33}$ by using of 16 Supernova-associated GRBs (SN/GRBs). In addition, they also utilized 14
kilonova-associated GRBs (KN/GRBs) to build a power-law relation of $L_{\gamma,p}\propto \tau^{-2.17\pm0.57}$.

Two leading types of central engines, i.e., (1) a hyper-accreting
black hole (BH) and (2) a rapidly spinning, strongly magnetized neutron
star (NS), have been proposed to power the GRB outflows
\citep[e.g.][]{Usov1992,Popham1999,Zhang2001}. Long-lasting emissions of
GRBs in the forms of extended emissions and X-ray afterglows, are
crucial to reveal the physical origin of the central engines
\citep{ZhangB2006,Norris2006,Rowlinson2013,Lu2015,Lu2018,Lu2020,Kisaka2017,Sharma2021}.
Based on the dipole spin-down model, \cite{Lu2015} confirmed the
hypothesis of the magnetar central engine model for short GRBs
(SGRBs) with an ``internal plateau'' followed by a very rapid
decay. In view of the energetics of GRBs, the central engines of
LGRBs with energies larger than 10$^{52}$ erg are preferentially
identified as black holes \citep{Sharma2021}. Nevertheless, the
unusual observations of low-luminosity GRBs (LLGRBs) suggest some
different emission processes, such as a less-collimated outflow, a
chocked jet, or even spherical ejecta moving at mildly
relativistic speeds \citep{Nakar2012,Bromberg2011}. Their
intrinsic features such as energetics, temperatures, and
timescales are explained very well by shock breakout emissions
\citep{Kulkarni1998,Nakar2012,Izzo2019}.

Several nearby LLGRBs such as GRB 980425/SN 1998bw,
GRB 031203/SN 2003lw, GRB 060218/SN 2006aj, GRB 100316D/SN 2010bh,
GRB 171205A/SN 2017iuk, GRB 190829A/SN 2019oyu, are identified to be
associated with SNe
\citep[e.g.][]{Galama1998,Malesani2004,Starling2011,Pian2006,Campana2006,Terreran2019}.
A systematic study of the time lags in the prompt emissions of
both the SN/GRBs and KN/GRBs \citep{Li2023} and the similarity
between SN/GRB 171205A and KN/GRB 170817A motivate us to undertake
further analysis in more detail. In fact, the properties of GRB
171205A were found to be different from other energetic GRBs and
other LLGRBs in many respects
\citep[e.g.][]{DElia2018,Wang2018,Izzo2019,Zhangxl2022}.
Therefore, it is necessary for us to go further to compare this
burst with other nearby LLGRBs systematically in this study. This
paper is organized as follows. Data analysis and methods are given
in Section 2. The main results are presented in Section 3.
Finally, the discussion and conclusions are presented in Section 4
and Section 5, respectively.

\section{Observations and data preparation} \label{sec:observations}
\subsection{Basic properties}
On 2017 December 5 at 07:20:43.9 UT, the Swift Burst Alert
Telescope (Swift/BAT) triggered and located GRB 171205A
\citep{DElia2017}, whose duration is long as $T_{\rm 90}$ = 190.5 $\pm$ 33.9 s as measured from T $-$ 26.2 to T $+$ 164.3 s
\citep{Izzo2017} and the time-averaged spectrum is best fit by
a power-law function with a photon index of $\alpha_{\rm PL} = -1.99 ^{+0.59}_{-0.82}$ while
the fluence and the peak photon flux in the 15 - 350 keV band are $6.78 ^{+0.99}_{-0.89} \times 10^{-6}$
erg cm$^{-2}$ and $1.02 ^{+0.29}_{-0.29}$ ph cm$^{-2}$ s$^{-1}$
\citep{Lien2016}. The afterglow observations of GRB 171205A
were carried out in multiple wavelengths from X-rays to radio
bands up to 1000 days
\citep[e.g.][]{Kennea2017,Butler2017,Izzo2017,Cobb2017,Choi2017,
Chandra2017a,Chandra2017b,DElia2018,Urata2019,
Leung2021,Maity2021}. Spectroscopic observations revealed that GRB
171205A is associated with a
 Type Ic SN 2017iuk
\citep{Izzo2017,Wang2018,DElia2018}. It is also found that GRB
171205A occurred in the outskirts of the bright spiral galaxy
2MASX J11093966-1235116 locating at $z = 0.037$, which is an
early-type (S0), high-mass, star-forming galaxy with a low
specific star formation rate and a low metallicity
\citep{Izzo2017,Wang2018}. Throughout this study, a flat
$\Lambda$CDM Universe with $\Omega_m = 0.286, \Omega_\Lambda =
0.714$ and $H_0 = 69.6$ km s$^{-1}$ Mpc$^{-1}$ is assumed
\citep{Bennett2014}.

\subsection{Data analysis}

To study the relation between luminosity and spectral lag for GRB 171205A, we first compare three different kinds of lags
including the cross correlation function (CCF) lag $\tau$ \citep{Band1997}, the centroid delay $\tau_{c}$ and the peak time delay $\tau_p$ \citep{Norris2000}
between channel 1 (15-25 keV) , channel 2 (25-50 keV), channel
3 (50-100 keV) and channel 4 (100-350 keV). It proves that the CCF method can provide the best lags with relatively smaller scatter (see Figure \ref{lagdistr} for a detail).
Consequently, the CCF lag will be calculated and applied for all GRBs in our sample herein. The mask-weighted light-curve data with
a 10 s resolution are taken from the Swift website
\footnote{\url{https://swift.gsfc.nasa.gov/results/batgrbcat}}
\citep{Lien2016}. We take two steps to obtain the more accurate
spectral lags. First, in order to obtain the smooth CCF curves, we
follow \cite{Zhangzb2006a,Zhangzb2006b,Zhangzb2008} to fit the
light-curve data using the ``KRL'' function of individual GRB
pulse and obtain four smooth light curve pulses
\citep{Lxj1,Lxj2,Lxj3}. Then, we perform the CCF analysis with the
Stingray PYTHON package
\citep{Huppenkothen2019a,Huppenkothen2019b}
\footnote{\url{https://docs.stingray.science/core.html}}.

Two distinct components could be observed following the prompt
emission of many GRBs. One is a high-level extended emission and
the other is a low-level plateau stage. \cite {Kisaka2017} used a
phenomenological formula to depict these two components, which
essentially synthesizes two functions (each has a constant flux
stage followed by a subsequent power-law decay). Considering that
the two power-law indexes could be different for the two
components, we modify their empirical formula as
\begin{equation} \label{eq2}
L_{\rm X,iso}(t)=L_{\rm EX,iso}(1+\frac{t}{T_{\rm EX}})^{\alpha_{1}}+L_{\rm PL,iso}(1+\frac{t}{T_{\rm PL}})^{\alpha_{2}},
\end{equation}
where $L_{\rm EX,iso}$, $L_{\rm PL,iso}$, $T_{\rm EX} $, and $
T_{\rm PL}$ are the luminosities and durations of the extended and
plateau emissions, and $\alpha_{1}$ and $\alpha_{2}$ are the
temporal indexes, respectively. Note that the extended
emission as emission with a timescale of $\lesssim 10^3$ s, some of which
are not detected by Swift/BAT. A longer timescale
component ($\gtrsim 10^3$ s) is defined as plateau emission \citep{Kisaka2017}.
The identification of the extended and plateau
emission is purely phenomenological \citep{Kisaka2017}.

\section{Results} \label{sec:results}

\begin{figure}
\begin{center}
\includegraphics[width=1\linewidth]{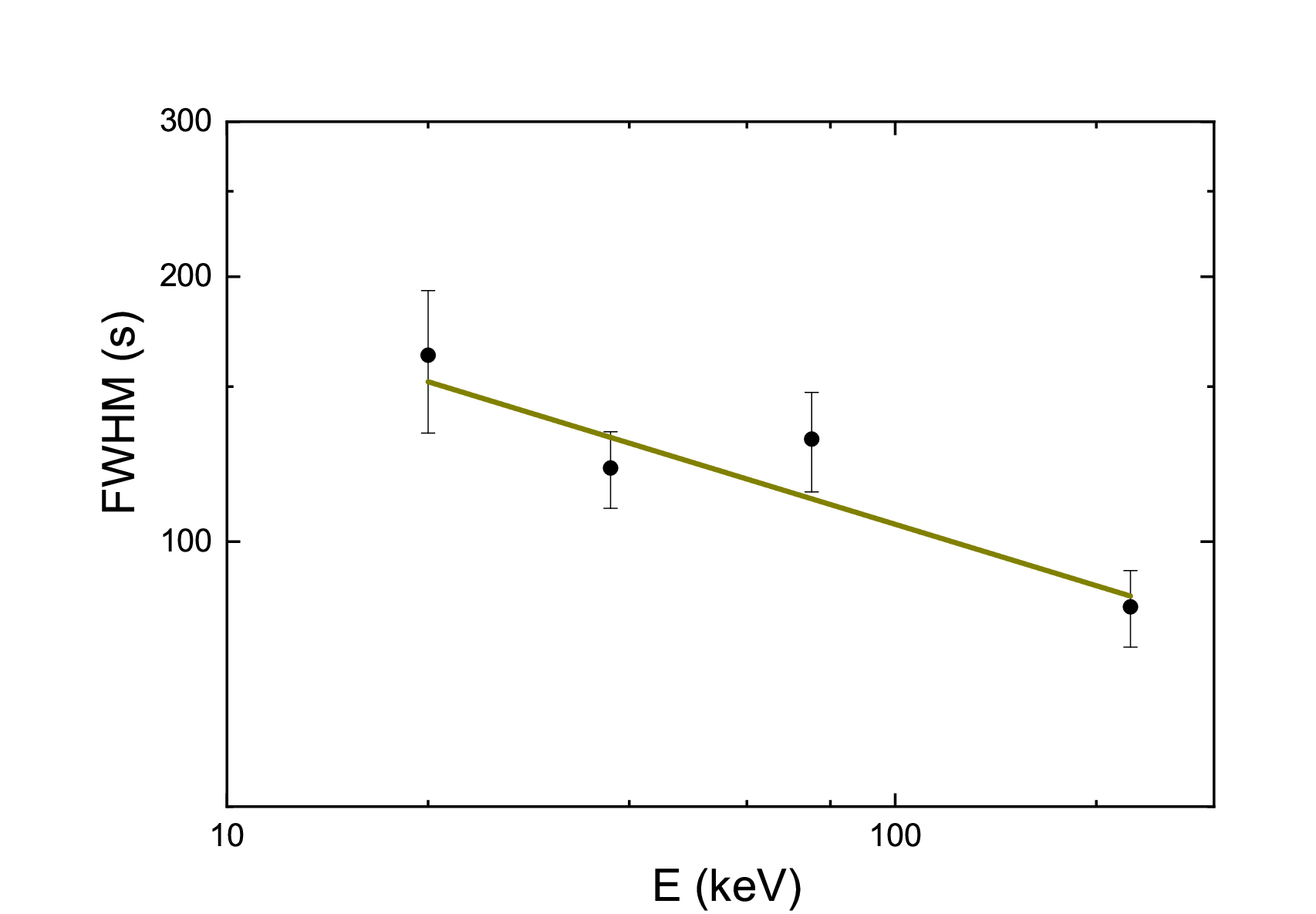}
\end{center}
\caption{The FWHM and the averaged photon energy of GRB 171205A in the four
     energy channels are anti-correlated. The solid line stands for the best
     power-law fit to the data.
\label{FE}}
\end{figure}

\subsection{Dependence of Pulse Width on Energy}

We now examine the dependence of the full width at half maximum
(FWHM) on the averaged photon energy ($E$). The FWHM versus photon
energy is plot in Figure~\ref{FE}. Here the errors of FWHM are
estimated by considering the error propagation, using the method
proposed in our previous study \citep{Zhangzb2006a}. Considering the fact that a larger bin size of light curve can improve the signal-to-noise level but also can smooth the pulse structures \citep{Lxj2}, we here use the observed light curves with a time resolution of 10s. It is feasible since GRB 171205A is a very long burst and it can be divided into sufficient bin numbers to ensure its essential temporal profiles.

It could be
seen that the negative power-law correlation, $FWHM \sim
E^\alpha$, still holds for GRB 171205A, with a Pearson coefficient
of $\rho$ = 0.92 and a chance probability of $P$ = 0.08. Our best
fit power-law index is $\alpha= -0.24\pm0.07$, which is coincident with those observed in GRB 980425 ($\alpha= -0.20\pm0.04$) and GRB 060218
($\alpha= -0.31\pm0.03$) \citep {Liang2006, Zhangfw2008}, but is
significantly larger than the value of $-0.4$ for normal LGRBs
\citep{Fenimore1995, Norris1996}. In addition, we
find that the power-law index of GRB 171205A is marginally
consistent with the mean value of $\alpha= -0.32\pm0.03$ for
single-peaked BATSE SGRBs \citep{Lxj1} and $\alpha= -0.32\pm0.02$ for Swift
SGRBs \citep{Lxj2}. Physically, the energy-dependent burst duration is
related to the evolution of peak energy across the observing
band \citep{Campana2006,Uhm2018}. Considering the fact that most KN/GRBs are
short in duration, one can conclude that SN /GRB 171205A shares the partial
properties of KN/GRBs, which has also been illustrated on the plane of luminosity
versus spectral lag \citep{Li2023}. The overlapping features of SN- and KN-GRBs
add more complexity to the GRB classification. The best solution is to assort
GRBs with more physical parameters jointly\citep[e.g.][]{Zhangb2009}.

\subsection{Peak Luminosity versus Time Lag}

\begin{figure}
\begin{center}
\includegraphics[width=1\linewidth]{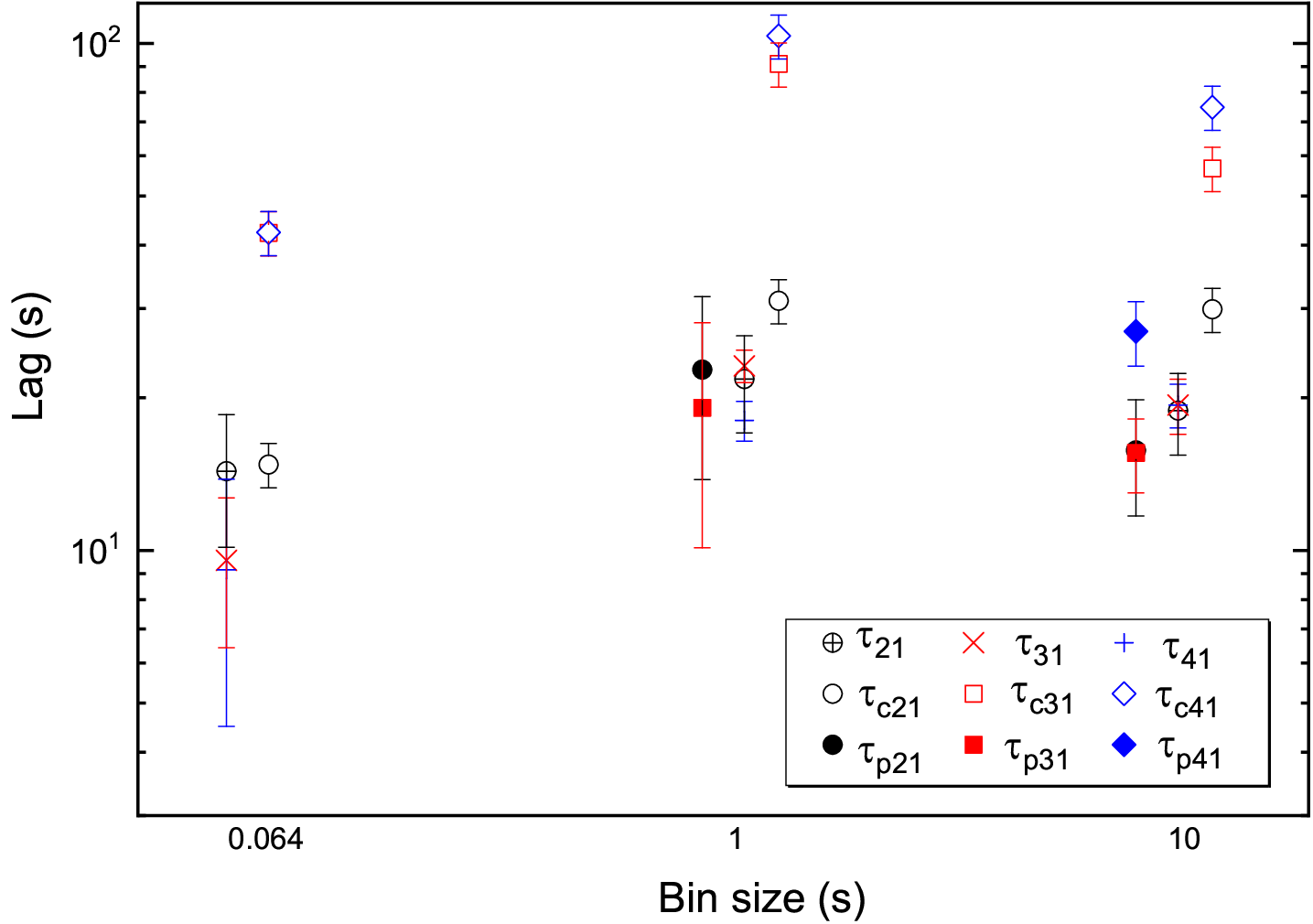}
\end{center}
\caption{The CCF lags (crosses), the peak lags (filled symbols) and the centroid lags (empty symbols) measured between different energy channels in time bins of 64 ms, 1 s and 10 s, respectively. \label{lagdistr}}
\end{figure}

To check whether GRB 171205A matches the $L_{\gamma,p}-\tau$ relation, one needs to precisely measure the time lags of light curves across different energy channels.
The CCF lag was adopted more frequent than either the peak lag or the centroid lag since these basic techniques were proposed in the past \citep{Band1997,Norris2000}.
Three types of lags between distinct energy channels in diverse time resolutions are compared for GRB 171205A in Figure~\ref{lagdistr}, where
it can be found that the CCF lags with a bin size of 10 s have relatively smaller errors and the centroid lags are larger than the others. Therefore, the CCF lag $\tau_{31}$ between energy channels 1 and 3 is chosen to compare GRB 171205A with other SN/GRBs and KN/GRBs in the plot of $L_{\gamma,p}$ versus $\tau_{31}$. The three types of lags are listed in Table \ref{Tab11}.

\begin{figure}
\begin{center}
\includegraphics[width=1\linewidth]{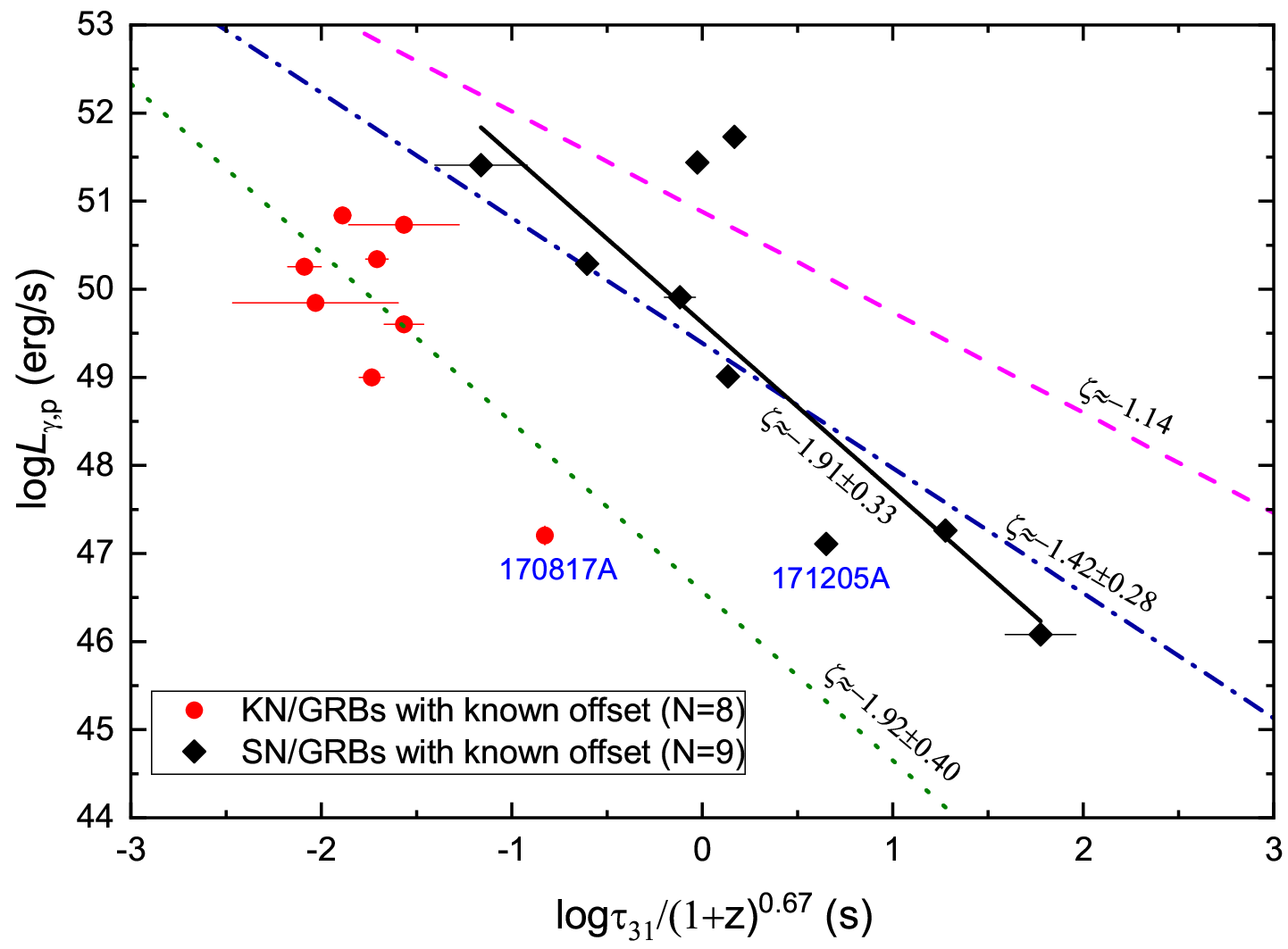}
\end{center}
\caption{The best fitting results of the $L_{\rm \gamma,p} - \tau_{31}$ relations for 9 SN/GRBs (solid line) with measured offsets in this work, for 16 SN/GRBs (dash-dotted line) and 14 KN/GRBs (dotted line) with known redshifts in \citet{Li2023}. The $L_{\rm \gamma,p} - \tau_{31}$ relation built with 7 LGRBs unassociated with SNe \citep{Norris2000} is also shown by the dashed line. \label{lag}}
\end{figure}

Figure~\ref{lag} is plotted to examine whether SN/GRBs and KN/GRBs with known offsets in our sample obey the power-law relations of $L_{\rm
\gamma,p}\propto\tau_{31}^{\varsigma}$, in which $L_{\rm \gamma,p}$ is calculated with
$L_{\rm \gamma,p}$ = 4$\pi D_{\rm L}^2 P_{\rm \gamma,bolo}$. Here
$D_{\rm L}$ is the luminosity distance, $P_{\rm \gamma,bolo}$ =
$P_{\rm\gamma}K_{\rm c}$ is the bolometric peak flux, and $K_{\rm
c}$ is the K-correction factor \citep{Zhangzb2018a,Zhangxl2020}.
The lags have been corrected by the factor $(1+z)^{-1}$ to compensate for
the cosmological time dilation. The spectral lag is proportional to the pulse width with a factor of (1+z) for the cosmological time dilation and a factor of $(1+z)^{-0.33}$ for the frequency shift \citep{Lxj1,Lxj2}. As a result, the ratio of the intrinsic time lag to the observed one is $(1+z)^{-0.67}$ \citep[see also][]{Gehrels2006}.
The errors of lag are calculated by considering the error propagation process as done in our previous study \citep{Zhangzb2006a}. We find that the lags of the SN/GRBs are longer than those of the KN/GRBs, which can be understood if the time lags are related to the burst durations through the emission radius and Lorentz factor \citep{Zhangb2009}. In particular, we find the new luminosity relation is
\begin{equation}
logL_{\rm \gamma,p}=(49.62\pm0.35)-(1.91\pm0.33)log\tau_{31},
\end{equation}
with a Pearson coefficient of $\rho$ = 0.91 and a chance probability of $P$ = 7.11$\times10^{-4}$ for the SN/GRBs, which is consistent with those based on a large sample of SN/KN-connected GRBs \citep{Li2023} but differ from the one given by \cite{Norris2000} for six normal LGRBs. For the KN/GRBs, it can be seen that most lags are nearly
zero, which is very close to previous
results for short GRBs \citep[e.g.][]{Norris2006,Zhangzb2006b,Bernardini2015}. However, it is hard to obtain a firm conclusion due to the limited sample of KN/GRBs with measured offset.
In addition, we caution that GRB 171205A exhibits a time lag close to the averaged one of normal LGRBs \cite[see e.g.][]{Norris2000, Zhangzb2006a,Zhangzb2006b} and it lies near the fitting line of SN/GRBs, which further strengthens its association with a supernova.

Since most bight GRBs are found to have narrower pulses and smaller lags \citep{Band1997,Norris2000}, it is straightforward to expect the power-law relation between luminosity and spectral lag as shown in Eq. (2). Because short GRBs have zero lags due to very fast variability \citep{Zhangzb2006b}, the luminosity-lag/time correlation could be insignificant among them, while the power-law relation is quite significant for long GRBs \citep{Norris2000,Zhangzb2006a}.

\begin{table*}
\centering
\small
\caption{The detailed values of lags under different bin sizes. \label{Tab11}}
\begin{tabular}{lcccccccccccccc}
\hline
Bin size & $\tau_{\rm p21}$$^a$&$\tau_{\rm p31}$$^a$ & $\tau_{\rm p41}$$^a$ & $\tau_{\rm c21}$$^b$&$\tau_{\rm c31}$$^b$&$\tau_{\rm c41}$$^b$&$\tau_{\rm 21}$$^c$&$\tau_{\rm 31}$$^c$&$\tau_{\rm 41}$$^c$\\
\hline
\hline
0.064 s& $--$$^d$ & $--$$^d$  & $--$$^d$ &14.77$\pm$1.48&42.31$\pm$4.23&42.37$\pm$4.24&9.16$\pm$4.66&9.56$\pm$3.13&14.33$\pm$4.19\\
1 s&22.72$\pm$8.93&19.12$\pm$9.00&$--$$^d$ &31.09$\pm$3.11&91.10$\pm$9.11&103.40$\pm$10.34&18.05$\pm$1.62&23.12$\pm$1.70&21.78$\pm$4.73\\ 	
10 s&15.76$\pm$4.06 &15.58$\pm$2.59 &27.02$\pm$3.93&29.89$\pm$2.99&56.68$\pm$5.67&74.81$\pm$7.48&19.69$\pm$1.91&19.36$\pm$2.41&18.88$\pm$3.46\\
\hline

\end{tabular}
\begin{list}{}{}
      \item Note:
      \item $a-$  the peak lags.
      \item $b-$  the centroid lags. We calculate the centroid as $t_{\rm centoid} = \sum I(t)t\Delta t /\sum I(t)\Delta t$
, where $\Delta t$ is the time bin of the observed data and $I(t)$ is the pulse intensity \citep{Zhangfw2008}. Note that the inferred errors of the centroid lags are large and we just take 10\% of the centroid lags as an error estimation in our calculations.
      \item $c-$  the CCF lags.
      \item $d-$  the signal in the corresponding bin size is too weak to be well fitted by a pulse function.\\
\end{list}
\end{table*}

\subsection{Extended Emission and Plateau Emission}
\begin{figure*}
\begin{center}
\includegraphics[width=1\linewidth]{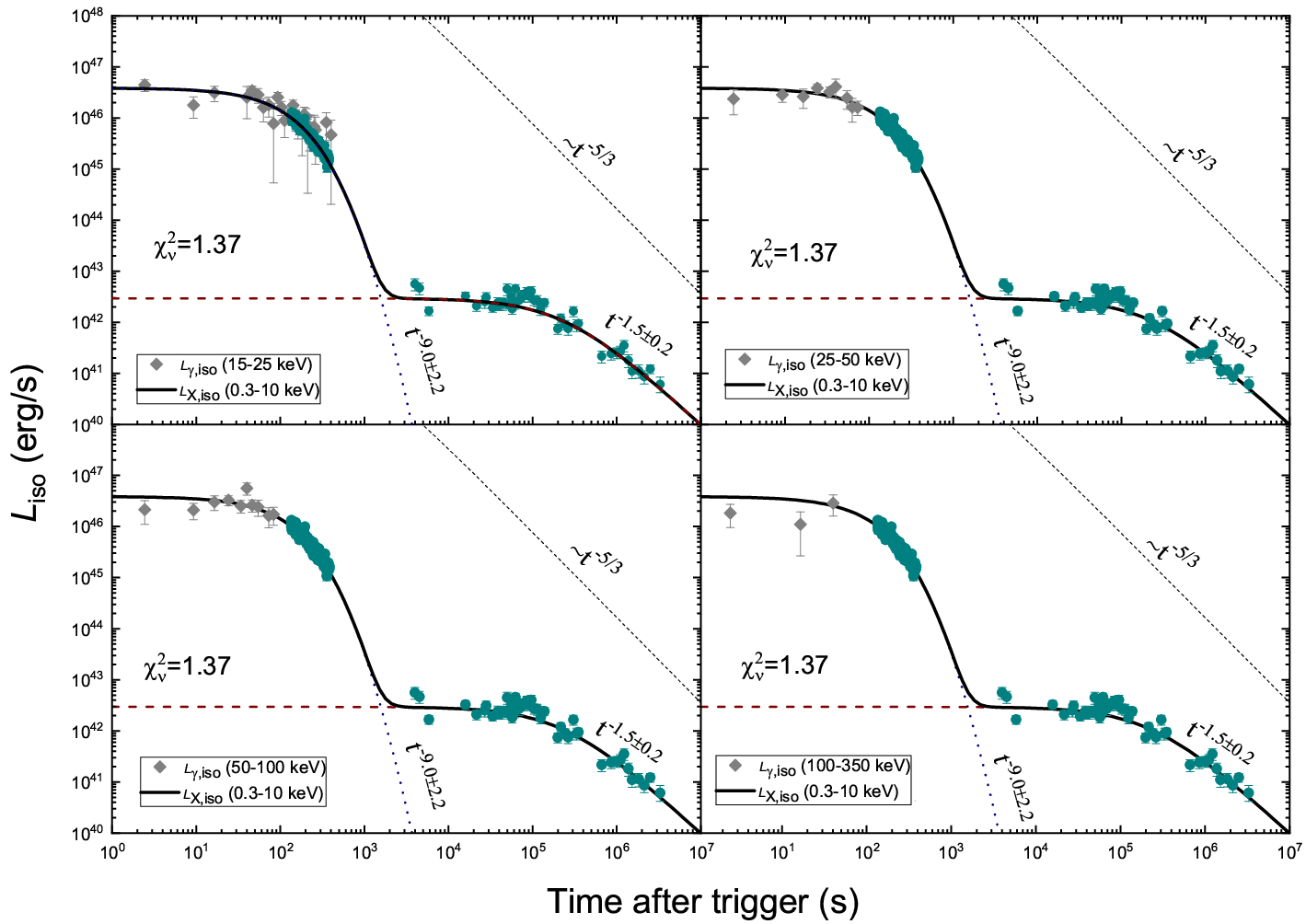}
\end{center}
\caption{The joint light curves of the prompt multi-wave band $\gamma$-ray emissions (gray diamonds) and the X-ray emissions
(green circles) of GRB 171205A. Note that the data points in BAT bands are chosen with high SNR (SNR $\geq$ 2). The solid line stands for the best
fit to the XRT data with Equation (\ref{eq2}). The dotted lines show the total luminosity of the Blandford-Znajek (BZ) jet launched by an evolving BH with a power-law timing index of $\alpha=-5/3$ \citep{Rosswog2007,Kisaka2015,Kisaka2017}. \label{X}}
\end{figure*}

\begin{figure*}
\begin{center}
\includegraphics[width=0.9\linewidth]{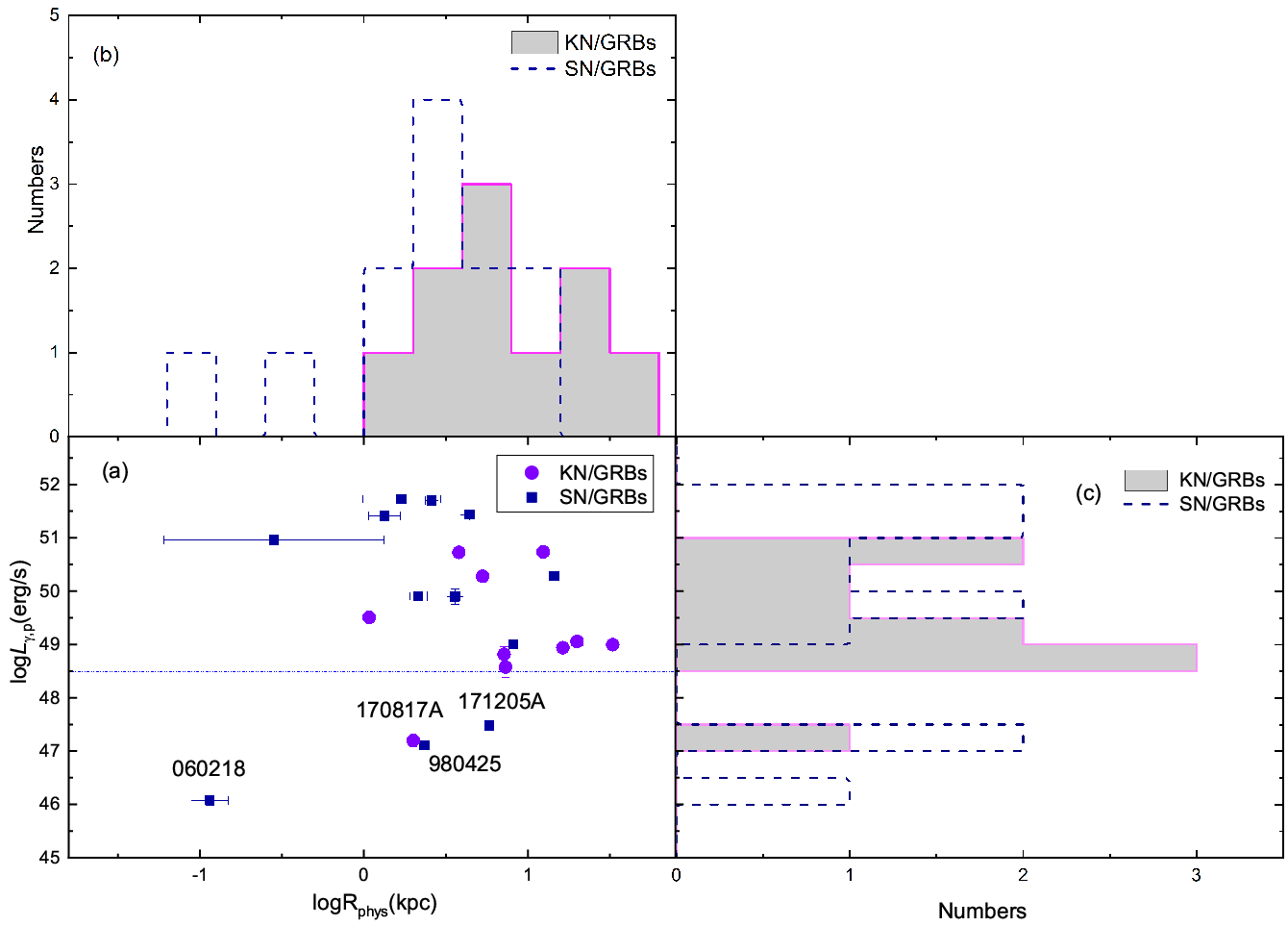}
\end{center}
\caption{One-dimensional (Panels b and c) and two-dimensional
   (Panel a) $L_{\rm \gamma,p}$ - $R_{\rm phys}$ distributions of 12
   SN/GRBs and 10 KN/GRBs. In Panel a, the long horizontal line
   corresponds to the upper limit of $L_{\rm \gamma,iso} \sim
   10^{48.5}$ erg s$^{-1}$ for LLGRBs derived by Cano et al. (2017).
\label{host2}}
\end{figure*}

\begin{figure}
\begin{center}
\includegraphics[width=1\linewidth]{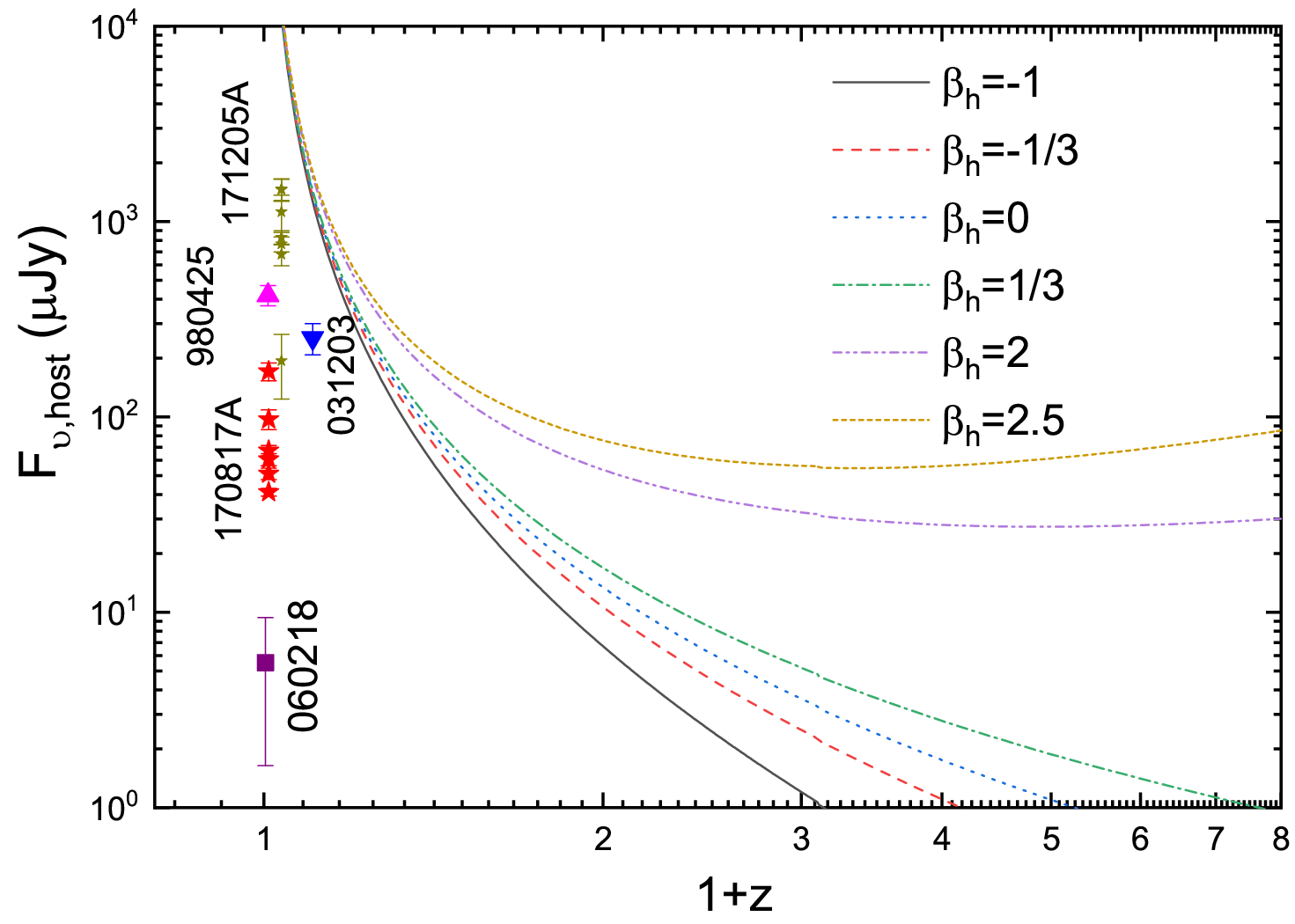}
\end{center}
\caption{Radio flux densities of GRB host galaxies versus their redshifts. The lines correspond to different spectrum indexes of
   $\beta_h = -1$ (solid), $-1/3$ (dashed), 0 (dotted), 1/3 (dash-dotted), 2 (dot-dot-dashed), and 2.5 (short dashed) in Zhang
   et al. (2018b). The $F_{\rm\nu, host}$ values of GRBs 060218, 980425, and 031203 are also taken from Zhang et al. (2018b). The
   $F_{\rm\nu, host}$ values of GRBs 171205A (yellow circle) and 170817A (red star) are estimated in this study, at the frequencies of 3.5 GHz and 7.1 GHz, respectively.} \label{host}
\end{figure}

\begin{figure*}
\begin{center}
\includegraphics[width=0.9\linewidth]{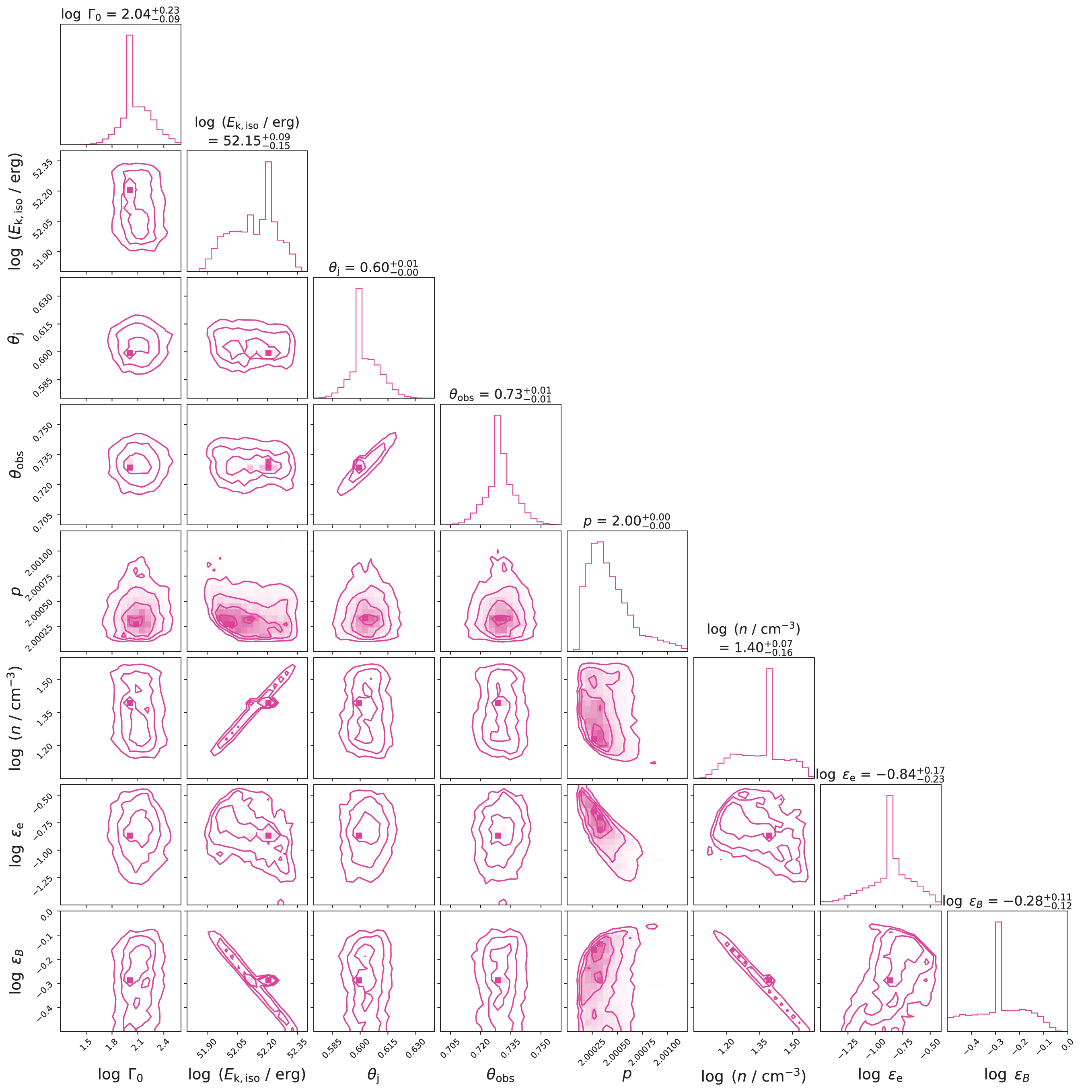}
\end{center}
\caption{Physical parameters derived by using a top-hat jet model ($1\sigma -3\sigma$ confidence levels) for GRB 171205A. The best fitting results are marked with $1\sigma$ uncertainties above the panel of their posterior distributions.
\label{un2}}
\end{figure*}
\begin{figure*}
\begin{center}
\includegraphics[width=1\linewidth]{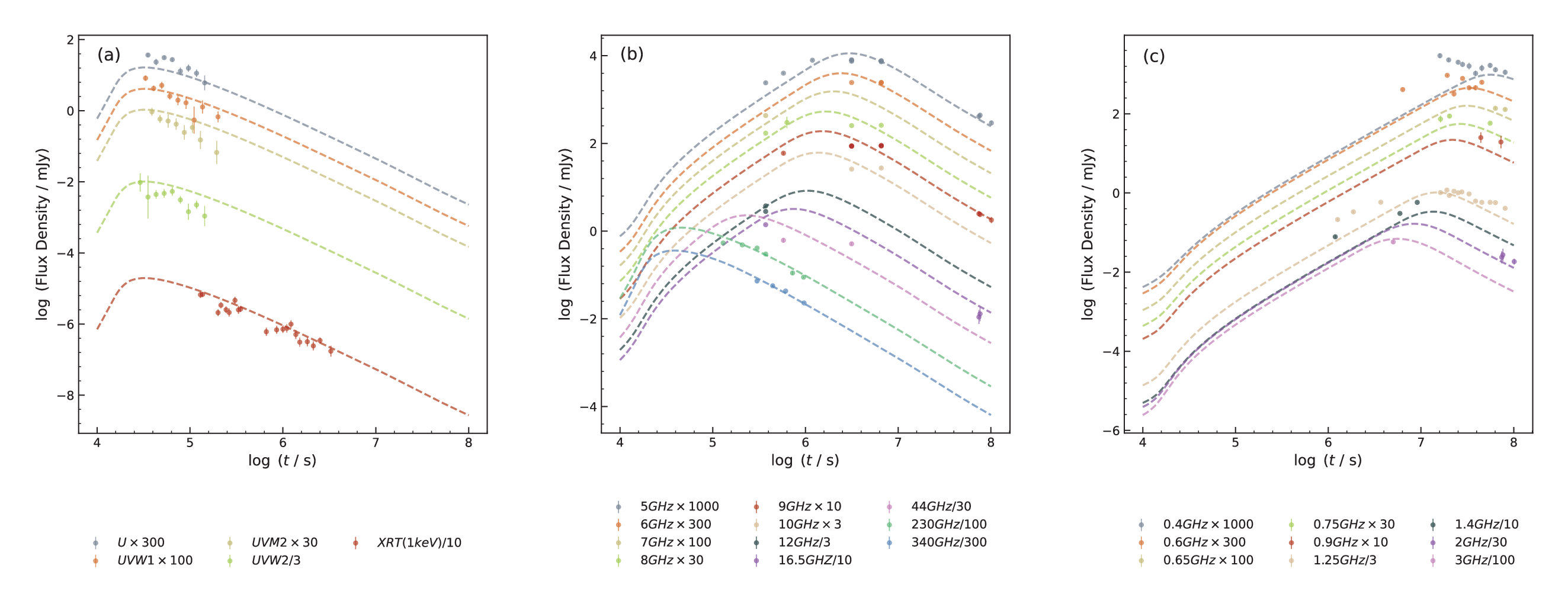}
\end{center}
\caption{Observed multi-wavelength afterglow of GRB 171205A and the best fitting lightcurves predicted by a top-hat jet model (dashed curves). The data of Swift/XRT/UVOT afterglow light curves in panel (a) are taken from \cite{DElia2018}. The radio afterglow data in panels (b) and (c) are obtained from \cite{Maity2021} and \cite{Leung2021}.  } \label{data3}
\end{figure*}

By jointly analyzing both the
Swift/BAT\footnote{\url{https://swift.gsfc.nasa.gov/results/batgrbcat}}
and
XRT\footnote{\url{https://www.swift.ac.uk/xrt_products/index.php}}
data, we now investigate the potential connection between the
extended emission and the plateau component. The X-ray luminosity
of GRB 171205A can be calculated as $L_{\rm X,iso}$ = 4$\pi D_{\rm
L}^2 F(1+z)^{-(2-\Gamma)}$ \citep{Tang2019,Xu2021}, where $\Gamma
= 1.63 $ is the photon spectral index taken from the Swift GRB
table\footnote{\url{https://swift.gsfc.nasa.gov/archive/grb_table.html/}}.
The prompt $\gamma$-ray light curves in the four BAT energy channels and the X-ray plateau emissions of GRB 171205A are presented in
Figure~\ref{X}, where we find that the Swift/BAT light curves in all energy bands perfectly bridge with the XRT afterglow.
This indicates that both $\gamma$-ray and early X-ray components should reflect the activities of the central engine
together and could share the same radiation mechanism.

Furthermore, we perform a temporal fit to the X-ray light curve of
GRB 171205A by adopting Equation \ref{eq2}. The
best-fit results are $L_{\rm EX,iso} = (3.86 \pm 0.11) \times
10^{46}$ erg s$^{-1}$, $L_{\rm PL,iso} = (2.95 \pm 0.24) \times
10^{42}$ erg s$^{-1}$, $T_{\rm EX}  = (840.03 \pm 241.66)$ s, $
T_{\rm PL} = (2.40 \pm 0.08) \times 10^{6}$ s, $\alpha_{\rm 1}$ = $-9.0\pm2.2$, and $\alpha_{\rm 2}$ = $-1.5\pm0.2$ with a reduced
chi-square of $\chi_\nu^2 \simeq 1.37$ (see Figure~\ref{X}).
The late X-ray plateau followed by a shallower decay could be mainly
contributed by the energy injection from a magntar \cite[e.g.][]{Zhang2001,Fan2006,Tang2019}, the jet viewed off-axis \citep{Beniamini2020} or the fall-back accretion process of black hole (BH) \citep{Yu2015}, which reflects the multiple
activities of the central engine. Of course, the possibility that the X-ray plateau is contributed
by a Supernova SN2017iuk cannot be fully ruled out \citep{DElia2018,
Li2023}. Interestingly, GRB 171205A favors a magnetar origin although it has
an external plateau in X-rays \citep{Wang2022}. The power-law index of $\alpha_{\rm 2}$ = $-1.5\pm0.2$ of the second
decay segment at late time (tens of days after trigger) is very close to the
expected value of $(4-3p)/2 \sim 1.6$ for an electronic spectral index of $p = 2.4$ in the
standard afterglow theory \citep{Gao2013},
which demonstrates that the late-time decay should be dominated by the
interaction of external shocks with circumburst material. This does not conflict with the magnetar origin that can be dominant in the early X-ray phase. Very recently, we analyzed those GRBs with internal X-ray plateaus and found that the prompt gamma-ray durations are tightly correlated with the plateau lasting times \citep{Du2023}, which indicates that the early X-ray emissions including plateaus should originate from some internal processes, such as the dissipation of magnetars.

\subsection{The host and offset}

For standard GRBs and high-luminosity GRBs, \cite{Zhangzb2018b}
found a similar redshift independence of the flux for host galaxies. Only three nearby LLGRBs
deviate from the relation. \cite{Li2015} statistically
investigated the relation between the host flux density ($F_{\rm
host}$) and the peak afterglow flux density ($F_{\rm o,peak}$) in
radio bands, and derived a useful tight correlation as $F_{\rm\nu,
host} = (b_1 + b_2 \nu )F_{\rm o,peak}$ for LLGRBs, with $b_1=0.27
\pm 0.02$ and $b_2 =-0.016 \pm 0.002$, where $\nu$ is the
observing frequency. Using this correlation, we can estimate the
$F_{\rm\nu, host}$ of GRB 171205A at frequencies with $F_{\rm
o,peak}$ available. Using the radio data provided in
\cite{Leung2021} and \cite{Maity2021}, we have calculated the
radio flux density of the host galaxy of GRB 171205A. The result
is illustrated in Figure~\ref{host} and is compared with several
other GRB hosts. It can be seen that the host spectral index ($\beta_h<-1$) of GRB 171205A is significantly less than the average spectral index $<\beta_h> \sim -0.75$ of spiral galaxies \citep{Condon1992}, which is similar to
other nearby LLGRBs, but differs from the high-luminosity or standard GRBs. In addition, we find that GRB 170817A as a
peculiar short GRB associated with both gravitational wave and
kilonova is located in the region with a lower spectral index of $\beta_h<-1$. It is worth noting that some selection effects including the instrumental threshold and the galaxy identification as emphasized by \cite{Zhangzb2018b} have been neglected in the work.

\begin{table*}
\centering
\small
\caption{Physical parameters of the SN/GRBs and KN/GRBs. \label{Tab1}}
\begin{tabular}{l c c c c c}
\hline
GRB & $T_{\rm 90}$&z & offset ($R_{\rm phys}$) & $logL_{\rm \gamma,p}$&Ref.\\
{}    & (s) & {} & (kpc)&(erg s$^{-1}$) &\\
\hline
KN/GRBs\\
\hline
170817A&$2.048$&0.009783&2.00$\pm$0.02&47.20$\pm$0.11&[3][8]\\
160821B&0.48&0.161&16.40$\pm$1.64$^e$&48.94$^{+0.05}_{-0.05}$$^c$&[3][10]\\
150101B&0.012&0.1343&7.30$\pm$0.05&48.58$^{+0.20}_{-0.19}$$^c$&[3][6]\\
130603B&0.176&0.3565&5.30$\pm$0.20&50.28$\pm$0.05&[2][5]\\
070809&1.28&0.2187&20.00$\pm$1.60&49.06$\pm$0.04&[2][5]\\
070714B&3.00&0.9224&12.40$\pm$0.90&50.74$\pm$0.02&[2][5]\\
061201&0.76&0.111&32.90$\pm$0.10&49.00$\pm$0.02&[1][2]\\
060614&109.104&0.125&1.08$\pm$0.04&49.51$\pm$0.02&[2][5]\\
060505&4.0&0.0889&7.16$\pm$0.06&48.82$^{+0.18}_{-0.15}$$^c$&[3][4]\\
050709&0.07&0.161&3.80$\pm$0.38$^e$&50.73$^{+0.05}_{-0.06}$&[3][5][8]\\
\hline
SN/GRBs\\
\hline
171205A&189.4&0.0368&5.81$\pm$0.58$^{d,}$$^e$&47.26$\pm$0.09&[2][11]\\
120422A&5.35&0.28&8.17$\pm$0.82$^e$&49.01$\pm$0.08&[2][5]\\
100621A&66.3&0.542&0.28$\pm$0.44 &50.96$\pm$0.02&[2][4]\\
091127&9.57&0.49&1.34$\pm$0.30 &51.41$\pm$0.02&[2][4]\\
090618&115.2&0.54&4.40$\pm$0.29 &51.44$\pm$0.01&[2][4]\\
090424&50.30&0.544&2.59$\pm$0.24 &51.71$\pm$0.03&[2][4]\\
081007&5.6&0.529&14.44$\pm$0.25 &50.29$\pm$0.05&[2][4]\\
080319B&147.0&0.937&1.70$\pm$0.93 &51.73$\pm$0.03&[2][4]\\
060729&120.0&0.54&2.15$\pm$0.27 &49.91$\pm$0.04&[2][4]\\
060218&2100.0&0.033&0.12$\pm$0.03 &46.08$\pm$0.09&[2][4][12]\\
050824&38&0.83&3.60$\pm$0.41 &49.90$\pm$0.14&[2][4]\\
980425&18.0&0.00866&2.34$\pm$0.01&47.11$\pm$0.02$^e$&[7][9]\\
\hline
\end{tabular}
\begin{list}{}{}
\item Ref. [1] \cite{Lu2015}; [2] \cite{Dainotti2020}; [3] \cite{Jin2021}; [4]\cite{Blanchard2016}; [5]\cite{Liye2016}; [6]\cite{Fong2016};
[7] \cite{Bloom2002}; [8]\cite{Wang2018}; [9] \cite{Norris2000}; [10] \cite{Zhangb2021}; [11] \cite{DElia2018}; [12] \cite{Campana2006};\\
$^c$ Calculated as $L_{\rm \gamma,p}$ = 4$\pi D_{\rm L}^2 P_{\rm \gamma,bolo}$.\\
$^d$ A projected angular offset of $R_{\rm ang} = 8$ arcsec is reported by \cite{DElia2018}.
     We calculate the projected physical offset as $R_{\rm phys} = D_{\rm L}/(1+z)^2 R_{\rm ang}$ \citep{Bloom2002}.\\
$^e$ Taking 10\% of the measured parameter as the error estimation.\\
\end{list}
\end{table*}
\begin{table*}
\centering
\small
\caption{The best-fit results of GRB 171205A for a Top-hat jet model. \label{Tab2}}
\begin{tabular}{l c c c c c c c}
\hline
$log \Gamma$ & $log (E_{\rm k,iso}/erg)$ & $\theta_j$ & $\theta_{obs}$ & $p$ &$log (n/cm^3)$ &$log\epsilon_e$&$log\epsilon_B$.\\
\hline
$2.04 ^{+0.22}_{-0.09}$ & $52.15 ^{+0.09}_{-0.15}$ & $0.60 ^{+0.01}_{-0.01}$&$0.73 ^{+0.01}_{-0.01}$ & $2.00 ^{+0.00}_{-0.00}$& $1.40 ^{+0.07}_{-0.16}$ &$-0.84 ^{+0.17}_{-0.23}$ &$-0.28 ^{+0.11}_{-0.12}$ \\
\hline
\end{tabular}
\end{table*}
The offsets of GRBs in their host galaxies can help reveal the
populations of unusual GRB progenitors \citep{Bloom2002}.
\cite{Dainotti2020} reported 22 SN/GRBs and 8 KN/GRBs.
\cite{Jin2021} provided another sample including 9 KN/GRBs. Here,
we choose these SN/KN-associated GRBs with projected physical
offset ($R_{\rm phys}$) or angular offset ($R_{\rm ang}$)
available to examine their similarity in depth. In total, we have
a sample consisting of 12 SN/GRBs and 10 KN/GRBs. Table \ref{Tab1}
lists the key parameters of these SN/GRBs and KN/GRBs and relevant
references. The overall distribution of offsets can provide a robust clue to the nature of the progenitors \citep{Bloom2002}. \cite{Zhang2017} have checked the possible correlations between the luminosities of short GRBs with/without extended emission and their offsets to examine the underlying physical origins. Here, similar studies are made to diagnose the possible difference of underlying physical origins between SN/GRBs and KN/GRBs. Figure~\ref{host2} illustrates their one-dimensional
and two-dimensional distributions on the $L_{\rm
\gamma,p} - R_{\rm phys}$ plane. We see that the offsets of
SN/GRBs tend to be smaller than those of KN/GRBs. This result is
consistent with the fact that SGRBs typically have a larger offset
than that of LGRBs \citep[e.g.][]{Fong2010,Fong2013}.

We have performed the Kolmogorov-Smirnov (K-S) test to analyze the
difference between the two distributions of KN/GRBs and SN/GRBs.
In panel (a), we get $D = 0.64$ from the K-S test, with a p-value
of 0.01. Adopting the critical value of $D_{\rm\alpha} = 0.74$ at
a significance level of $\alpha = 0.005$, we conclude that the
distribution of SN/GRBs is different from that of KN/GRBs. It is
further noticed that four LLGRBs, i.e. GRBs 171205A, 170817A,
060218, and 980425, are generally located in the lower section and
obviously deviate from the other GRBs due to their very low
luminosities. Additionally, we find that there is a moderate correlation
between the isotropic prompt luminosity and the
offset, with the Pearson correlation coefficient being $\rho =
-0.44$ $(-0.30)$ and a chance probability of $P = 0.23$ $(0.43)$ for
SN/GRBs (KN/GRBs) after excluding the four LLGRBs. It is similar
to the result of \cite{Zhang2017}. Since the offset distributions of SN/GRBs and KN/GRBs with longer and shorter durations, are obviously distinct as illustrated by \cite{Troja2008}, their luminosities should be related with offsets. However, due to the limited
number of LLGRBs, no general conclusion on the correlation between luminosity and offset can be drawn currently.

\subsection{Modelling multi-wavelength afterglows  }

We adopt an overall dynamic evolution and radiation process of jetted GRB
ejecta model \citep{Huang1999,Huang2000,Huang2006} to fit the multi-wavelength
afterglow of GRB 171205A on condition that a simple top-hat jet model is assumed.
We use the Markov Chain Monte Carlo (MCMC) algorithm to get the best fitting
result for the multi-wavelength GRB afterglow. The corresponding corner plot of some typical parameters is shown in
Figure ~\ref{un2} and Table 3, in which the jet half-opening angle and the
viewing angle are found to be $\thicksim$ 34.4 and 41.8 degrees, respectively,
confirming that GRB 171205A was viewed off-axis \citep[see e.g.][]{Maity2021,Kumar2022}.
Consequently, the large $\theta_j$ should correspond to the angular size of a cocoon
rather than the jet core \citep{Lxj1,Maity2021}. In addition, we present the observed
data of multi-wavelength afterglows and the best fitting light curves in Figure ~\ref{data3}.

The Karl G. Jansky Very Large Array Sky Survey (VLASS) revealed that the spectral luminosity of GRB 171205A lies between normal long GRBs and SNe with H-poor prompt spectrum \citep{Stroh2021}. In addition, \cite{Arabsalmani2022} presented a detailed study on the distribution and kinematics of atomic hydrogen in the host galaxy of GRB 171205A through the HI 21cm emission line observation with the JVLA. Its unusual HI features indicate that GRB 171205A could be ignited under extreme conditions with rare dynamics. Here, we find that GRB 171205A is located close to other SN/GRBs in the plane of peak luminosities versus spectral lags, which is consistent with the result of all SN/KN-associated GRBs \citep{Li2023}. The host galaxy spectrum of GRB 171205A, like other nearby LLGRBs, has a spectral index lower than -1. Meanwhile, we notice in the plot of $L_{\rm \gamma,iso}$ against $R_{\rm phys}$ that GRB 171205A is located between SN/GRBs and KN/GRBs. However, the result is somewhat ambiguous owing to a limited number of KN/GRBs with measured offsets.

\section{Discussion } \label{sec:discussion}

The central engine of GRBs may be a magnetar, especially those GRBs with
a plateau component in the X-ray afterglow \citep{XuHuang2012,Tang2019}.
The total rotational energy of a millisecond magnetar can be written
as $ E_{\rm rot} = I \Omega_0^2/2 \simeq 2 \times10^{52} {\rm erg}
M_{1.4} R^2_6 P_{0,-3}^{-2}$, where \emph{I} is the moment of inertia,
$P_0 = 2\pi/\Omega_0$ is the initial spin period. $M$ and $R$ are the
mass and radius of the NS. The magnetar spins down due to magnetic
dipole radiation, and the spin-down luminosity evolves with time as
$L_{\rm EM}(t)=L_{\rm em,0}\left(1+t/\tau_{\rm c,em}\right)^{-2}$,
where $L_{\rm em,0} \simeq 1.0 \times 10^{\rm 49} B_{\rm p,15}^{\rm 2}
P_{\rm 0,-3}^{\rm -4} R_{\rm 6}^{\rm 6}$ erg s$^{\rm -1}$ is the
initial kinetic luminosity and $\tau_{\rm c,em} \simeq 2.05 \times
10^{\rm 3} $ s $I_{\rm 45} B_{\rm p,15}^{\rm -2} P_{\rm 0,-3}^{\rm 2}
R_{\rm6}^{\rm-6}$ is the characteristic spin-down time scale \citep{Lu2015,Lu2018}.
To diagnose the center engine of GRB 171205A, we use the Swift X-ray data to constrain the initial spin period and the dipolar surface magnetic field to test whether the results match the spin-up line predictions for typical neutron star \citep{Stratta2018}.
In this way, we constrain the initial spin period as $P_{\rm 0} \sim (117.6 \pm 0.4)$
ms, with the dipolar surface magnetic field being $B_{\rm p} \sim (3.73 \pm 0.33)
\times 10^{\rm 15}$ G. The magnetic field deviates from the bounding of the
$B_{\rm p} $-$ P_{\rm 0}$ parameters corresponding to the range of mass accretion
rates
$10^{\rm -4}$ $M_\odot$ $s^{\rm -1}$ $<$ \.{M} $<$ $0.1M_\odot$ $s^{\rm -1}$
\citep{Stratta2018}. Note that $M = 1.35 M_{\odot}$ and $R = 11.9$ km \citep{Deibel2013,Lattimer2016,Most2018}
have been adopted in our calculations.


In contrast to those normal GRBs produced from an ultra-relativistic jet driven by a compact
central engine, low-luminosity GRBs may be powered by shock breakouts
\citep[e.g.][]{Kulkarni1998,Nakar2012}. For example, \cite{Starling2012} argued
that two low-luminosity GRBs (060218 and 100316D) with thermal spectrum and emitting radius
much smaller than those of the normal energetic GRBs can be interpreted by the shock
breakout model. However, GRB 171205A as a typical low-luminosity burst
has a thermal spectral component but with a temperature close
to that of the thermal component in an energetic SN/GRB 101219B \citep{DElia2018}. This hints that
both low- and high-luminosity SN/GRBs may have the thermal spectrum universally.

\cite{Izzo2019} studied the multi-epoch spectrum of GRB171205A/SN2017iuk and argued that the high speed emission
features should originate from a mildly relativistic hot cocoon generated due to
the breakout of an ultra-relativistic jet \citep[see also][]{Suzuki2022}. The geometric feature of outflow is quite similar to that of the KN/GRB 170817A
on basis of multiple-facility radio observations \citep{Mooleyi2018a,Mooleyi2018b}. Interestingly, we found that the special structure of
a relativistic jet surrounded by a non-relativistic cocoon of GRB 170817A-like events can also be distinguished by the temporal evolution of prompt $\gamma$-ray light curves \citep{Lxj1}.
Subsequently, \cite{Maity2021} utilized the upgraded Giant Metre-wave Radio Telescope (uGMRT) to monitor the late
radio afterglow ($\sim1000$ days after burst) of GRB 171205A in a frequency range of 250--1450 MHz and further inferred GRB 171205A to originate from an off-axis jet enveloped
by a wide cocoon. \cite{Kumar2022} used the latest XRT data to constrain the jet half-opening angle to be $\theta_j > 51.3$ degrees. For such a large jet half-opening angle, the X-ray afterglow under off-axis condition is expected to peak dozens (or even hundreds) of days after the GRB detection and the X-ray flux would be very low. However, its X-ray afterglow peaks within $\sim$ 1 day, which resembles other GRB X-ray light-curves viewed on-axis.

In general, GRBs can be physically classified as short-hard (type I)
and long-soft (type II) groups according to the multiple classification
standards \citep{Zhangb2009}. The types I and II bursts are respectively
produced by the compact star merger and core-collapse processes. We notice
from \cite{Li2023} that almost all SN/GRBs (except GRB 200826A) have
durations longer than 2s while most KN/GRBs are short ones with $T_{90}<2$s.
There are four long KN/GRBs (050724, 060614A, 070714B and 080503) challenging
the traditional classification scheme on basis of
durations \citep{Kouveliotou1993,Zhangzb2008}. This means that a fraction of
long GRBs should originate from compact binary collisions instead of
core-collapse processes. For example, a recent striking long GRB 211211A
associated with kilonova provides a compelling evidence of a compact binary
merger origin \cite[e.g.][]{ Yang2022,Troja2022}. On the other hand, \cite{Li2023}
systematically compared the temporal and spectral properties of 53 SN/GRBs and
15 KN/GRBs and found a heavy overlap in the plots of luminosity-lag, Amati relation
and plateau duration-luminosity between both types of GRBs. In practice, most GRBs
do not have an SN or KN detected, but they still could be SN/GRBs or
KN/GRBs. In these cases, the SN or KN signals are not detectable because
either the signals are too faint or they are buried beneath the brighter GRB afterglow.
As a nearby SN/GRB, GRB 171205A exhibits both differences and similarities
with normal LGRBs, nearby LLGRBs and SGRBs, indicating that the SN/LLGRB
populations might be more complex than what we thought before. Hopefully,
our statistical results can shed new light on the nature of GRB 171205A
and provide some useful clues for further investigations in the field.

\section{Conclusions}

We find that the pulse width and energy of GRB 171205A have a
power-law relation with an index of $-0.24\pm0.07 $, which is on average
larger than those of normal long GRBs. The early X-rays and gamma-rays
may reflect the activities of the central engine, while the late X-rays
should be resulted from the interaction of the external shock
with the circumburst material. Using the 9 SN/GRBs with measured offset,
we found that $L_{\gamma,p}\propto\tau^{-1.91}$, which is different
from $L_{\gamma,p}\propto\tau^{-1.15}$ derived by \cite{Norris2000} for
normal long GRBs. The MCMC method is engaged to fit the multi-wavelength
afterglow of GRB 171205A. The jet half-opening angle and the viewing angle
are found to be $\thicksim$ 34.4 and 41.8 degrees, respectively,
confirming the off-axis geometry of this event.

\section*{Acknowledgements}

We thank the anonymous referee for very constructive comments and
suggestions that led to an overall improvement of this study.
This work was supported by the National Natural Science Foundation of China (No. U2031118),
the Youth Innovations and Talents Project of Shandong Provincial Colleges and Universities
(Grant No. 201909118) and the Natural Science Foundations (ZR2018MA030, ZR2023MA049, XKJJC201901).
YFH is supported by the National Key R\&D Program of China (2021YFA0718500),
by National SKA Program of China (No. 2020SKA0120300), and by the National Natural Science
Foundation of China (Grant Nos. 12233002, 12041306). YFH also acknowledges the support from
the Xinjiang Tianchi Program.


\listofchanges
\end{document}